\documentclass[prb,aps,showpacs,floatfix]{revtex4}

\usepackage{amsmath,amssymb}
\usepackage{graphicx}
\usepackage{version}

\newcommand{\bB}{{\bf B}}
\newcommand{\bQ}{{\bf Q}}
\newcommand{\bk}{{\bf k}}
\newcommand{\bK}{{\bf K}}

\newcommand{\br}{{\bf r}}
\newcommand{\bp}{{\bf p}}
\newcommand{\bq}{{\bf q}}

\newcommand{\bA}{{\bf A}}
\newcommand{\ba}{{\bf a}}

\newcommand{\bb}{{\bf b}}

\newcommand{\beqa}{\begin{eqnarray}}
\newcommand{\eeqa}{\end{eqnarray}}

\begin{document}

\title{Wigner crystal and bubble phases in graphene in quantum Hall regime }
\author{C.-H. Zhang}
\affiliation{Department of Physics, Indiana University-Purdue
University Indianapolis, Indianapolis, Indiana 46202, USA}
\author{Yogesh N. Joglekar}
\affiliation{Department of Physics, Indiana University-Purdue
University Indianapolis, Indianapolis, Indiana 46202, USA}
\date{\today}
\begin{abstract}
Graphene, a single free-standing sheet of graphite with honeycomb
lattice structure, is a semimetal with carriers that have linear dispersion.
A consequence of this dispersion is the absence of Wigner crystallization
in graphene, since the kinetic and potential energies both scale identically
with density of carriers. We study the ground state of graphene in the
presence of strong magnetic field focusing on states with broken
translational symmetry. Our mean-field calculations show that at integer
fillings a uniform state is preferred whereas at non-integer fillings,
Wigner crystal states (with broken translational symmetry) have lower
energy. We obtain the phase diagram of the system. We find that
it is qualitatively similar to that of quantum Hall systems in
semiconductor heterostructures. Our analysis predicts that non-uniform
states, including Wigner crystal state, will occur in graphene in the
presence of a magnetic field and will lead to anisotropic transport in high
Landau levels.
\end{abstract}

\maketitle


\section{Introduction}
\label{sec:intro}

Two dimensional electron gas (2DEG), realized either in
semiconductor heterostructures or by sprinkling electrons on Helium
surface, has been extensively studied over the past few decades. It
is known that such a system undergoes phase transition from a
uniform state to a state with spontaneously broken translational
symmetry, a Wigner crystal, when the carrier density is sufficiently
low\ \cite{wigner1934} or the temperature is sufficiently
high.~\cite{grimes1979} In case of a 2DEG in semiconductors, this
state occurs when the gain from lowering the potential energy by
localization outweighs the kinetic energy cost associated with the
localization {\it provided that the dispersion of carriers is
well-described by an effective mass $m^{*}$}, $E_{k}=\hbar^2
k^2/2m^{*}$. This transition has been theoretically investigated,\
\cite{tanatar1989} although unequivocal experimental evidence is
still lacking. The spectrum of the electron gas changes radically in
the presence of a strong magnetic field. The kinetic energy is
quantized and each Landau level (the manifold of eigenstates with a
given energy) has a macroscopic degeneracy. Since the kinetic energy
is fixed for a given Landau level, it is possible to vary the ratio
of potential energy and kinetic energy by changing the filling
factor within a given Landau level, and the system undergoes a
transition from a uniform state to states with broken translational
symmetry. Indeed, this transition has been extensively studied
theoretically.\ \cite{Das97} There is strong experimental evidence
that the ground state of such a system at filling factors $\nu<1/5$
is a Wigner crystal\ \cite{willett1988} and that at partial filling
factors in high Landau levels, the ground state is non-uniform.\
\cite{mikestripes} We remind the Reader that this evidence is
obtained from transport measurements and that a direct measurement
of spatial density modulation - crystalline structure - is
exceedingly difficult since the 2DEG is buried under a substrate.

In this paper, we focus on how these results change when the carries
in the 2D gas have a linear dispersion instead of the the usual
effective-mass quadratic dispersion. Graphene, a single sheet of
graphite with honeycomb lattice structure, is a realization of a
system with such carriers. It has the added advantage that such a 2D
gas of carriers is {\it not buried under a substrate} and is,
therefore, amenable to local probes that can investigate the
crystalline structure. Graphene is a semimetal in which the valance
and conduction bands touch at two inequivalent points $\bK$ and
$\bK'=-\bK$ (and four other points related by symmetry). In the
vicinity of these points (valleys), the band structure of carriers
is well described by $E_{\bk}=\pm\hbar v_G k$ where $\hbar v_G=5.8$
eV\AA\, is the characteristic velocity and $\bk$ is measured from
one of the six points at which the conduction and valance bands
touch.\ \cite{wallace1947,Novoselov05} Due to the linear dispersion
of carriers in graphene, potential and kinetic energies both scale
as $n^{3/2}$ with the carrier density $n$. Therefore, graphene does
not undergo Wigner crystallization in the absence of an external
magnetic field.\ \cite{hari}

In the presence of a magnetic field, however, the kinetic energy is
quantized and its ratio with the potential energy can be varied by
changing the filling factor. This raises the prospect of Wigner
crystal states in graphene. Here, we present a systematic mean-field
analysis of the ground state of graphene in the quantum Hall regime,
focusing on partial filling of the first few Landau levels. The plan
for the paper is as follows. In the next section, we set up the
low-energy Hamiltonian for graphene in the presence of magnetic
field and recall results for the single-particle spectrum. In Sec.\
\ref{sec:HFA}, we set up the formalism for Hartree-Fock (HF)
approximation and discuss its details. In Sec.\ \ref{sec:numerics},
we present the numerical results we obtain. We discuss the phase
diagram of graphene as a function of the partial filling factor and
the density profiles. We find that the results closely follow those
of non-uniform states in the conventional 2DEG. We end the paper
with a brief discussion and conclusions in section\ \ref{sec:end}.


\section{Graphene in Magnetic Field}
\label{sec:graphene}

The low-energy Hamiltonian for electrons in the $\bK$ valley is
given by\ \cite{Semenoff84,Haldane88}
\begin{equation}
\label{eq:dirac}
H_{\bK} =v_G\left(p_x\tau_x +p_y\tau^*_y\right),
\end{equation}
where $\tau_x$ and $\tau_y$ are Pauli matrices in the space consisting of
two lattice sites $A$ and $B$ within a single unit cell (The Hamiltonian for
the other valley is obtained by complex conjugation). The Hamiltonian in the
presence of an magnetic field is obtained by Peirels substitution
$\bp\rightarrow\bp-e\bA/c$. In a uniform magnetic field $\bB=B\hat{z}$,
generated by a vector potential $\bA=Bx\hat{y}$, the Hamiltonian
(\ref{eq:dirac}) becomes
\begin{equation}
\label{eq:diracgraphene}
H_\bK =\frac{\sqrt{2}\hbar v_{G}}{l_B}\left(\begin{array}{cc} 0 & c_k \\
c^{\dagger}_k & 0\end{array}\right),
\end{equation}
where $c_k=-i[l_B\partial_x+(x/l_B-kl_B)]/\sqrt{2}$ is the lowering
operator, $l_B=\sqrt{\hbar c/eB}$ is the magnetic length, and
$[c_k,c_k^{\dagger}]=1$. The eigenvalues of this Hamiltonian are given by
$E_n=\pm\hbar v_G\sqrt{2|n|}/l_B$ and the corresponding
eigenfunctions are given by
\begin{equation}
\langle\br|\bK,nk\rangle =\frac{1}{\sqrt{2L_y}}e^{iky}
\left[\begin{array}{c}\mbox{sgn}(n)\varphi_{|n|-1}(x-kl_B^2)\\
\varphi_{|n|}(x-kl_B^2)\end{array}\right]
\label{eq:efnonzero}
\end{equation}
for $n\ne0$  and
\begin{equation}
\langle\br|\bK,0k\rangle =\frac{1}{\sqrt{L_y}}e^{iky}
\left[\begin{array}{c}0 \\\varphi_{0}(x-kl_B^2)
\end{array}\right]
\label{eq:efzero}
\end{equation}
for $n=0$. Here, $\varphi_n(x)$ are the simple harmonic oscillator
eigenfunctions defined by $\varphi_n(x)=\langle x|(c^{\dagger}_k)^n
|0\rangle/\sqrt{n!}$ and $L_y$ is the sample length in
$y$-direction. (In the following, we will use units such that
$l_B=1$). Thus, $n=0$ eigenfunctions in graphene are identical to
$n=0$ states in a conventional 2DEG, whereas for $n\neq 0$, the
eigenfunctions of graphene are an admixture of wavefunctions on the
$A$ and $B$ lattice sites.\ \cite{peres2006} Therefore, we expect
that graphene, like conventional 2DEG at partial filling factors,
will support non-uniform (Wigner crystal) states.

In the following, we denote crystals with one electron per unit
cell, $N_e=1$ as Wigner crystals, and those with $N_e\ge 2$ per unit
cell as bubble crystals.\ \cite{Koulakov96,goerbig04} We also
consider modulated stripe states that can be described by oblique,
rectangular, or centered rectangular lattices.\
\cite{Cote00,Ettouhami06} We call these states anisotropic Wigner
crystals, and the ones having a triangular or square lattice
structure as isotropic Wigner crystals.


\section{Hartree-Fock Hamiltonian}
\label{sec:HFA}

The microscopic Hamiltonian for carriers in graphene consists of the
kinetic energy and Coulomb repulsion. We use a pseudospin notation to denote
the valley index: $\sigma=+$ corresponds to the $\bK$ valley and
$\sigma=-$ corresponds to the $\bK'=-\bK$ valley. In the single-particle
basis (\ref{eq:efnonzero},\ref{eq:efzero}) the Hamiltonian is
\begin{align}
\label{eq:H_int} \hat{H} &
=N_{\phi}\sum_{n\sigma}(E_n-\mu)\hat{\rho}^{\sigma,\sigma}_n(0) +
\frac{N_{\phi}}{4\pi l_B^2}\sum_{\bq,\{\sigma n\}} V(\bq)
\mathcal{F}_{n_1,n_4}(\bq)\mathcal{F}_{n_2,n_3}(-\bq)
\hat{\rho}^{\sigma_1,\sigma_1}_{n_1,n_4}(-\bq)
\hat{\rho}^{\sigma_2,\sigma_2}_{n_2,n_3}(\bq),
\end{align}
where $N_{\phi}=A/(2\pi l_B^2)$ is the number of flux quanta in
the area $A$ of the sample, $\mu$ is the chemical
potential, $V(\bq)=2\pi e^2/\epsilon q$ is the Coulomb interaction
and $\epsilon$ is the dielectric constant of graphene ($\epsilon\sim
3$), $\hat{\rho}^{\sigma,\sigma^\prime}_{n,n^\prime}(\bq)$ is the
density matrix element defined as
\begin{align}
\hat{\rho}^{\sigma,\sigma^\prime}_{n,n^\prime}(\bq)
=\frac{1}{N_{\phi}}\sum_{kk^\prime}
e^{-\frac{i}{2}q_x(k+k^\prime)l_B^2}c^\dagger_{\sigma nk}c_{\sigma
n^\prime k^\prime} \delta_{k,k^\prime+q_y}
\end{align}
with $c^\dagger_{\sigma nk}$ ($c_{\sigma nk}$) the creation
(annihilation) operator of the electrons,
$\mathcal{F}_{nn^\prime}(\bq)$ is the form factor
\begin{align}
{\cal F}_{n_1,n_2}(\bq)&=\delta_{n_1,0}\delta_{n_2,0}
F_{n_1,n_2}(\bq)+\frac{1}{\sqrt{2}}\delta_{n_1n_2,0}\delta_{n_1+n_2\ne0}
F_{n_1,n_2}(\bq)
\nonumber\\
&+\frac12\theta(|n_1|)\theta(|n_2|)\left[F_{|n_1|,|n_2|}(\bq)
+\mathrm{sgn}(n_1n_2)F_{|n_1|-1,|n_2|-1}(\bq)\right],
\end{align}
where $\theta(x)$ is the heavyside function and $\mathrm{sgn}(x)$ returns
the sign of its argument. This form factor is a linear combination of 
form factors for wave functions on the two inequivalent lattice sites,
\begin{equation}
F_{n_1\geq n_2}(\bq)=\sqrt{\frac{|n_2|!}{|n_1|!}}\left[\frac{(-q_y+iq_x)}
{\sqrt{2}}\right]^{(|n_1|-|n_2|)}L_{|n_2|}^{|n_1|-|n_2|}\left(\frac{q^2}{2}
\right)e^{-q^2/4}
\end{equation}
and $F_{n_1\leq n_2}(\bq)=F_{n_2,n_1}(-\bq)^{*}$. Here $L_{n}^m(x)$
is the generalized Laguerre polynomial. To obtain the Hamiltonian
(\ref{eq:H_int}), we have ignored the interaction terms that scatter
electrons from one valley to another and are exponentially and
algebraically small in $a/l_B$ where $a$ is the lattice constant of
graphene ($a\sim$ 5 \AA, $l_B\sim$100 \AA).\ \cite{goerbig06}

The derivation of the HF Hamiltonian from Eq.\ (\ref{eq:H_int}) is
straightforward and has been discussed extensively in the
literature.\ \cite{Cote91,Cote92,Chen92} When inter-Landau level
transitions are ignored, the Hartree-Fock Hamiltonian for a single Landau 
level index is given by
\begin{equation}
\label{eq:hf}
H_{HF}=\frac{N_\phi e^2}{\epsilon l_B}\sum_{\sigma\bQ}\left\{ \left[
\frac{(E_n-\mu)}{e^2/\epsilon l_B}\delta_{\bQ,0}+ H_{n}(\bQ)-X_{n}^{\sigma,
\sigma}(\bQ)\right]
\hat{\rho}_{n}^{\sigma,\sigma}(\bQ)-X_{n}^{\sigma,\bar{\sigma}}(\bQ)
\hat{\rho}_{n}^{\bar{\sigma},\sigma}(\bQ)\right\}
\end{equation}
where $\bQ$ is a reciprocal lattice vector of the Wigner crystal and
$\bar{\sigma}=-\sigma$. The dimensionless Hartree and Fock potentials are
given by
\begin{align}
H_{n}(\bQ)& =\frac{e^{-Q^2/2}}{Q}|{\cal F}_{n,n}(\bQ)|^2\rho_n(-\bQ)
(1-\delta_{\bQ,0}),\\
X_{n}^{\sigma,\sigma^\prime}(\bQ) &=\int_0^{\infty}\,dx
e^{-x^2/2}|{\cal F}_{n,n}(x)|^2 J_0(xQ)\rho_{n}^{\sigma\sigma'}(-\bQ)
\end{align}
where
$\rho_n^{\sigma,\sigma'}(\bQ)=\langle\hat{\rho}^{\sigma,\sigma'}_{n,n}
(\bQ)\rangle$ are determined self-consistently from the 
Hamiltonian (\ref{eq:hf}) and $\rho_n(\bQ)=\sum_{\sigma=\pm}
\rho_n^{\sigma,\sigma}(\bQ)$ is the total density at wavevector
$\bQ$.

The density matrix $\rho^{\sigma,\sigma'}_{n}(\bQ)$ is determined from the
equal-time limit ($\tau\rightarrow 0^{-}$) of the single-particle Green's
function
\begin{align}
G^{\sigma'\sigma}_{n}(k_1,k_2;\tau)=-\langle\mbox{T}c_{nk_1\sigma}(\tau)
c^\dagger_{nk_2\sigma'}(0)\rangle.
\end{align}
We define the Fourier transform of $G^{\sigma,\sigma'}_{n}(k_1,k_2;\tau)$ as
\begin{align}
G^{\sigma,\sigma'}_{n}(\bQ,i\omega_m)=\frac{1}{N_\phi}
\sum_{k_1k_2}\int _0^{\beta} d\tau e^{-iQ_x(k_1+k_2)/2+i\omega_m\tau}
\delta_{k_2,k_1-Q_y} G^{\sigma,\sigma'}_{n}(k_1,k_2;\tau)
\end{align}
where $\beta=1/k_BT$ is the inverse temperature. The equation of
motion for $G^{\sigma,\sigma'}_{n}(\bQ,i\omega_m)$ is given by\
\cite{Cote92,Chen92}
\begin{align}
\label{eq:G}
\left[\begin{array}{c} \delta_{\bQ,0}\\
0\end{array}\right] =(i\omega_m+\mu)\left[\begin{array}{c}
G^{+,+}_n(\bQ,i\omega_m)\\
G^{-,+}_n(\bQ,i\omega_m)\end{array}\right]
-\sum_{\bQ'}\left[\begin{array}{cc}
\Sigma^{+,+}_n(\bQ,\bQ') &
\Sigma^{+,-}_n(\bQ,\bQ')\\
\Sigma^{-,+}_n(\bQ,\bQ') &
\Sigma^{-,-}_n(\bQ,\bQ')\end{array}\right]
\left[\begin{array}{c} G^{+,+}_n(\bQ',i\omega_m)\\
G^{-,+}_n(\bQ',i\omega_m) \end{array}\right]
\end{align}
where the Hartree and exchange self-energy terms are given by
\begin{align}
\label{eq:sigma}
\Sigma^{\sigma,\sigma}_n(\bQ,\bQ')& =\left[H_n(\bQ'-\bQ)-X_n^{\sigma,\sigma}
(\bQ'-\bQ)\right]
\exp\left(\frac{i}{2}\bQ\times\bQ'\cdot\hat{z}\right),\\
\Sigma^{\sigma,\bar{\sigma}}_n(\bQ,\bQ')&=
-X^{\sigma,\bar{\sigma}}_n(\bQ'-\bQ)\exp\left(\frac{i}{2}\bQ\times\bQ'\cdot
\hat{z}\right)
\end{align}
and $\bQ\times\bQ'\cdot{z}=Q_xQ'_y-Q_yQ'_x$. In order to solve Eq.\
(\ref{eq:G}), we diagonalize the self-energy matrix
\begin{align}
\label{eq:diag}
\sum_{\bQ'}\left[\begin{array}{cc}
\Sigma^{+,+}_n(\bQ,\bQ') &
\Sigma^{+,-}_n(\bQ,\bQ')\\
\Sigma^{-,+}_n(\bQ,\bQ') &
\Sigma^{-,-}_n(\bQ,\bQ')\end{array}\right]
\left[\begin{array}{c} V_j(\bQ')\\
U_j(\bQ') \end{array}\right] =\omega_j
\left[\begin{array}{c} V_j(\bQ)\\
U_j(\bQ) \end{array}\right]
\end{align}
in a basis with a specific lattice structure, where $(V_j,U_j)$ is the
$j$-th eigenvector and $\omega_j$ is its corresponding eigenvalue. Using
these eigenvectors and eigenvalues, one can calculate the density matrix
\begin{equation}
\label{eq:rhoq}
\left[\begin{array}{cc}
\rho^{+,+}_n(\bQ) & \rho^{+,-}_n(\bQ)\\
\rho^{-,+}_n(\bQ) & \rho^{-,-}_n(\bQ)
\end{array}\right] = \sum_j \left[V^*_j(0),U^*_j(0)\right]
\left[\begin{array}{c}
V_j(\bQ)
\\ U_j(\bQ)
\end{array}\right]f(\omega_j-\mu)
\end{equation}
where $f(\omega-\mu)$ is the Fermi-Dirac distribution function and the
chemical potential $\mu$ is determined by
\begin{equation}
\label{eq:rhoq0} \rho^{+,+}_n(0)=\sum_j
V_j(0)V^*_j(0)f(\omega_j-\mu)=\frac12\nu
\end{equation}
where $\nu$ is the partial filling factor in the Landau level $n$.
We solve the set of equations (\ref{eq:G}-\ref{eq:rhoq0})
self-consistently to obtain the resultant density matrix and the
ground state energy per particle for different lattice structures
The real-space density profile is then obtained using inverse
Fourier transform,
\begin{align}
\label{eq:rhor}
n(\br)=\frac{1}{2\pi l_B^2}\sum_{\sigma\bQ}\rho_{n}^{\sigma,\sigma}(\bQ)
{\cal F}_{n,n}
 (\bQ)e^{i\bQ\cdot\br}
\end{align}
For simplicity, in the numerical results for density profile, we use the
dimensionless density, $2\pi l_B^2 n(\br)$.


\section{Numerical Results}
\label{sec:numerics}

Now we turn to the numerical results. It follows from Eq.\
(\ref{eq:hf}) that a state with intervalley coherence always has
lower energy than a state without the coherence. Therefore, we only
focus on solutions with $\rho^{+,-}_n(\bQ)\neq 0$. Thus, the ground
state consists of electrons occupying the symmetric state (between
the two valleys) when the filling factor within a Landau level is
$\nu<1$. (Here, we have assumed spin-split Landau levels\
\cite{Zhang06} and ignored inter-Landau level transitions because
they do not qualitatively change our conclusions.)

We use a simplified oblique lattice with two primitive lattice
vectors\ \cite{brey2000} $\ba_1=(a,b/2), \ba_2=(0,b)$ (We perform
similar analysis with square, rectangular, and centered rectangular
lattices, and obtain results similar to those presented below). We
denote the ratio $\gamma=b/a$. Note that the triangular lattice
($\gamma=2/\sqrt{3}$) and quasi-striped states ($\gamma\rightarrow
0$) are its special cases. We do not consider purely one-dimensional
striped states because they are prone to density modulations along
the stripes.\ \cite{Cote00} The two lattice constants are determined
by the constraint that the unit cell contains $N_e$ electrons, and
are given by $a=l_B\sqrt{2\pi N_{e}/\nu\gamma}$ and $b=a\gamma$. The
reciprocal lattice basis vectors are
\begin{align}
\bb_1=\frac{2\pi}{a}(1,0),\
\bb_2=\frac{2\pi}{a}\left(-\frac{1}{2},\frac{1}{\gamma}\right),
\end{align}
and the reciprocal lattice vectors are given by $\bQ_{mn}=m\bb_1+n\bb_2$.
We determine the optimal lattice structure by obtaining the $\gamma$
($0<\gamma\le2/\sqrt{3}$) that minimizes the mean-field energy. We use an
external (infinitesimal) potential with the symmetry of the lattice to
generate the initial density matrix,
\begin{align}
\Sigma_{ex}(\bQ,\bQ')=-\exp\left[-\frac{Q_0^2}{2}-\frac{1}{4}
\left(Q^2+Q'^2\right)+
\frac{1}{2}\left(\bQ\cdot\bQ'+i\bQ\times\bQ'\cdot\hat{z}\right)\right]
\end{align}
where $Q_0=|\bQ_{11}|$. We adjust the number of basis states used to
calculate the density matrix, and verify that the zero-temperature
sum rule\ \cite{Cote92}
\begin{equation}
\sum_{\bQ}\left[|\rho^{+,+}_n(\bQ)|^2+|\rho^{+,-}_n(\bQ)|^2\right]
=\rho^{+,+}_n(0)
\end{equation}
is satisfied within an accuracy of $10^{-5}$.

\begin{figure}[tbh]
\includegraphics[width=8cm]{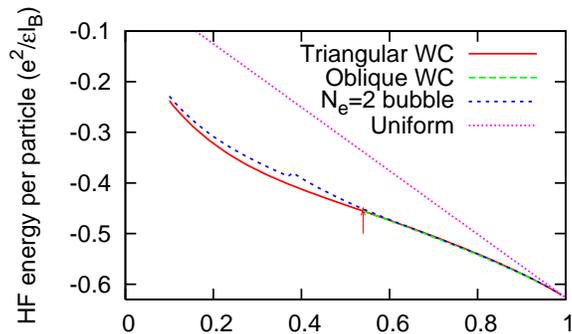}
\vspace{-8mm} \caption{(Color Online) Ground state energy per
particle (measured in units of $e^2/\epsilon l_B$) in graphene as a
function of partial filling $\nu$ in the $n=0$ Landau level. The
ground state is a triangular Wigner crystal for small $\nu$ and
becomes an anisotropic Wigner crystal as $\nu$ increases. The arrow
mark critical filling factors $\nu^*$ at which transition from the
triangular Wigner crystal to the anisotropic Wigner crystal takes
place. Note that for $n=0$, the bubble crystal always has higher
energy.} \label{fig:ehfnun0}
\end{figure}
\begin{figure}[tbh]
\includegraphics[width=13cm]{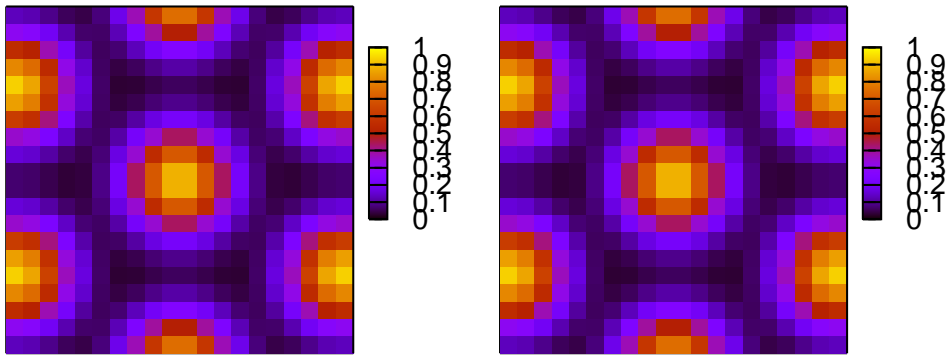}
\caption{(Color Online) A density plot of the ground state electron
density in graphene (left) and a bilayer system with $d=0$ (right)
in the $n=0$ Landau level with partial filling $\nu=0.25$. According
to the phase diagram in Fig.~\ref{fig:ehfnun0}, the ground state is
an isotropic triangular crystal. As expected, due to the equivalence
between the two systems in the $n=0$ Landau level, the two densities
are identical.} \label{fig:rhon0nu0.25}
\end{figure}
\begin{figure}[tbh]
\includegraphics[width=8cm]{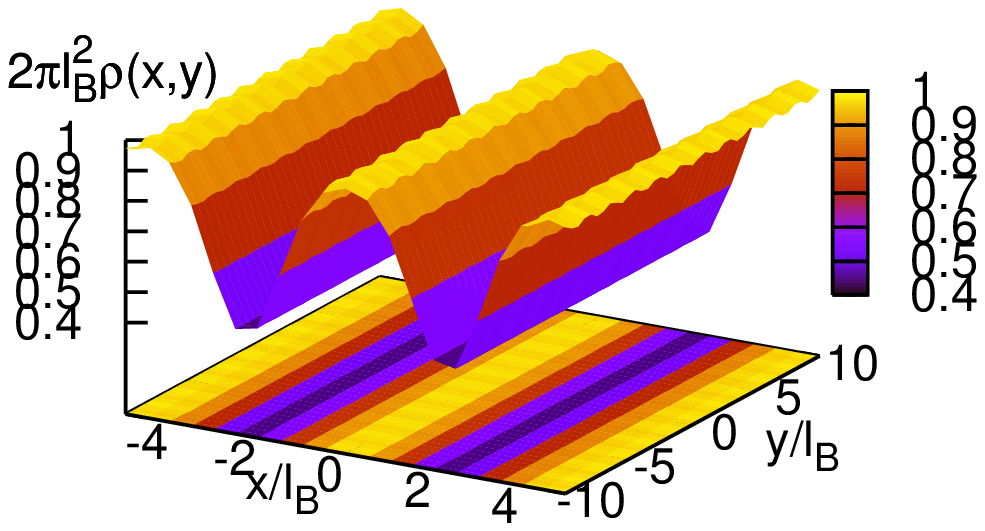}
\caption{(Color Online). Electron density profile for the
anisotropic Wigner crystal ground state at $\nu=0.75$ in the lowest
Landau level. Notice that the density modulation along the y-axis is
small and the system behaves essentially like a striped state.}
\label{fig:rhon0nu0.75}
\end{figure}

For $n=0$, the equivalence between single-particle wavefunctions for
graphene and conventional 2DEG, implies that at small filling
factors, the ground state is a triangular Wigner crystal.\
\cite{yangwc2001} Indeed, graphene, with its pseudospin (valley
index) can be mapped onto a bilayer system in which layer index is
the pseudospin, in the limit when the layer separation $d\rightarrow
0$. Our calculations reproduce the results for ground state energy
and lattice structure. Figure\ \ref{fig:ehfnun0} shows the
mean-field energy per particle (phase diagram) as a function of
partial filling factor for different lattice structures. We see that
for $\nu\leq 0.62$, the ground state is triangular Wigner crystal;
it becomes an anisotropic Wigner crystal for $\nu\ge0.62$. We find
that a bubble crystal with $N_e=3$ is identical, in energy, to the
triangular Wigner crystal, and the bubble crystals with $N_e\ge2$
have higher energies. Figure\ \ref{fig:rhon0nu0.25} shows the
real-space electron density profile for graphene at partial filling
$\nu=0.25$ and for a bilayer quantum Hall system at layer separation
$d=0$ at the same filling factor. We obtain, as expected, identical
density profiles with a triangular Wigner crystal. As the partial
filling $\nu$ is increased, the ground state of the system changes
to an anisotropic Wigner crystal (Figure\ \ref{fig:rhon0nu0.75}). We
find, in general, that the optimal value of anisotropy is high.
Therefore, the electron density resembles uniform stripe states and
the density modulation along the stripes is quite small (Figure\
\ref{fig:rhon0nu0.75}).

\begin{figure}[thb]
\includegraphics[width=10cm]{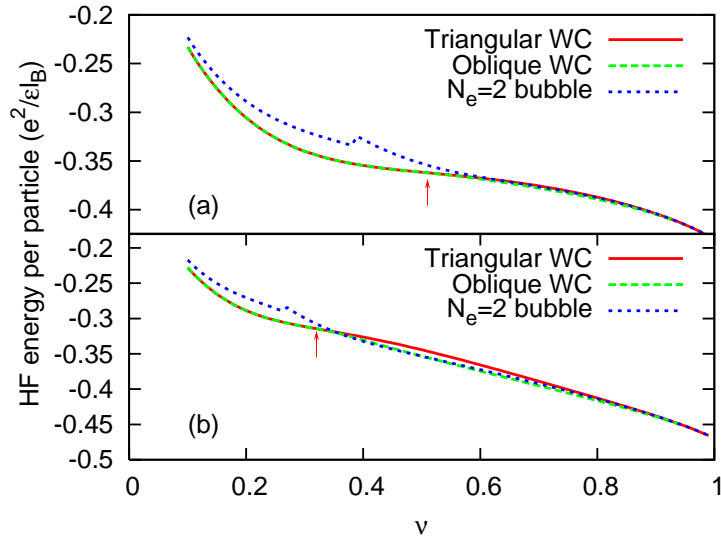}
\caption{(Color Online) Ground state energy per particle (measured
in units of $e^2/\epsilon l_B$) for different crystal structures in
the $n=1$ Landau level for (a) graphene, and (b) a bilayer system at
$d=0$. The arrows mark the critical filling factor $\nu^*$ at which
transition from the triangular Wigner crystal to anisotropic Wigner crystal
takes place.}
\label{fig:ehfnun1}
\end{figure}

\begin{figure}[thb]
\includegraphics[width=10cm]{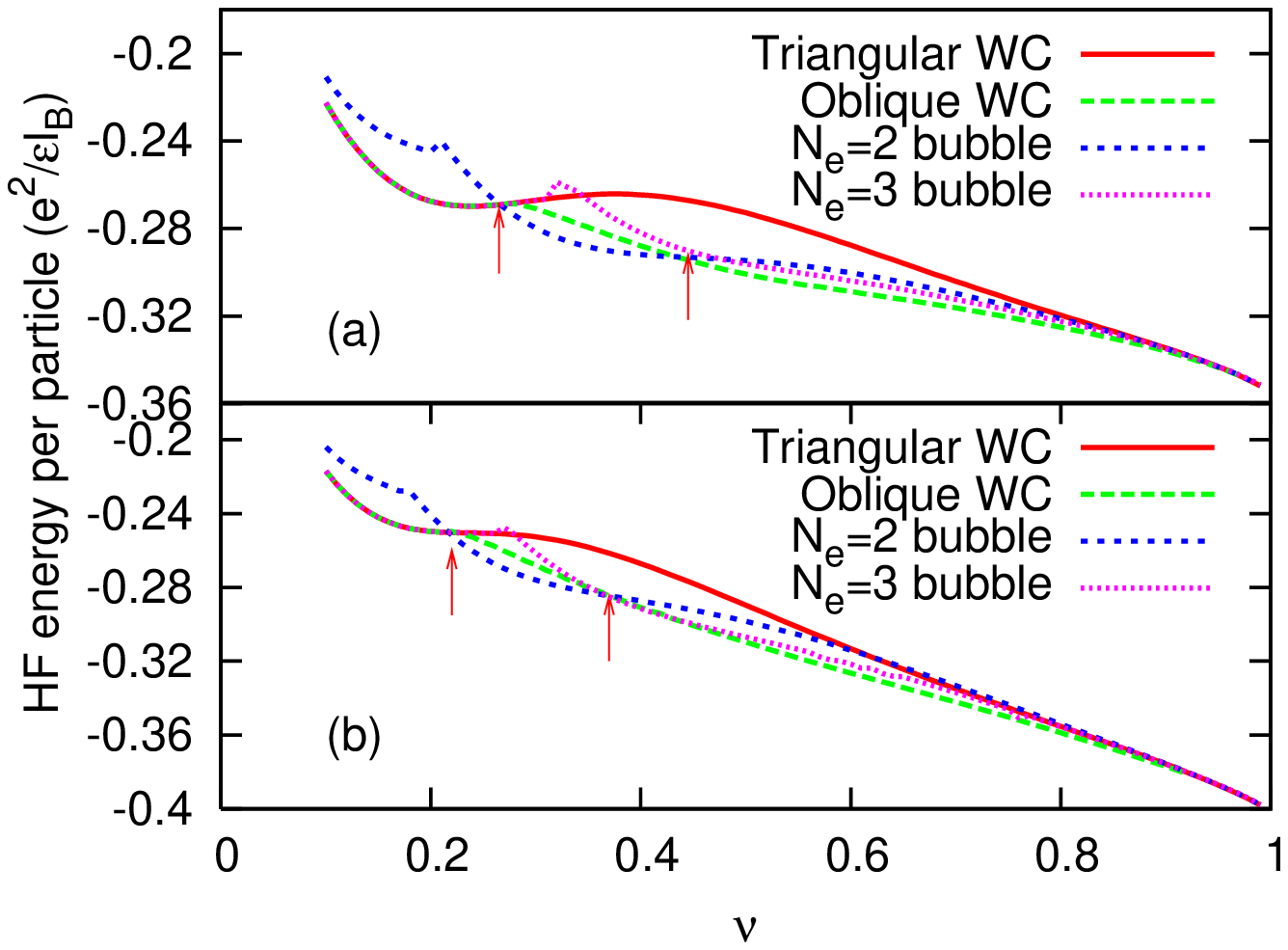}
\caption{(Color Online) Ground state energy per particle (measured
in units of $e^2/\epsilon l_B$) for different crystal structures in
the $n=2$ Landau level for (a) graphene, and (b) a bilayer system at
$d=0$. The arrows mark critical filling factors $\nu^*$ at which
transition from the triangular Wigner crystal to bubble crystal to
anisotropic Wigner crystal take place.}
\label{fig:ehfnun2}
\end{figure}

\begin{figure}[thb]
\includegraphics[width=10cm]{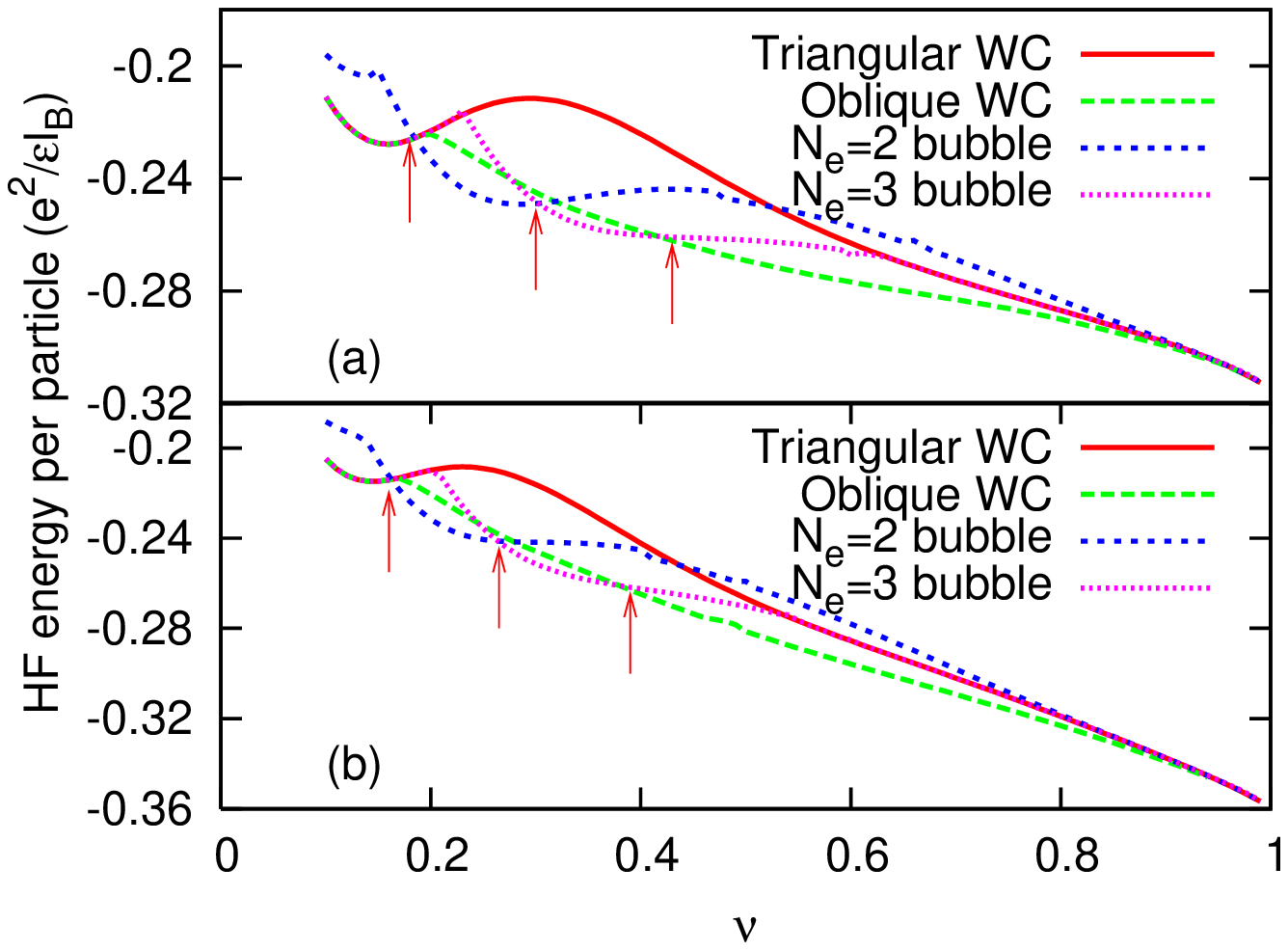}
\caption{(Color Online) Ground state energy per particle (measured
in units of $e^2/\epsilon l_B$) for different crystal structures in
the $n=3$ Landau level for (a) graphene, and (b) a bilayer system at
$d=0$. The arrows mark critical filling factors $\nu^*$ at which
transition from the triangular Wigner crystal to bubble crystal to
anisotropic Wigner crystal take place. }
\label{fig:ehfnun3}
\end{figure}

\begin{figure}[thb]
\includegraphics[width=10cm]{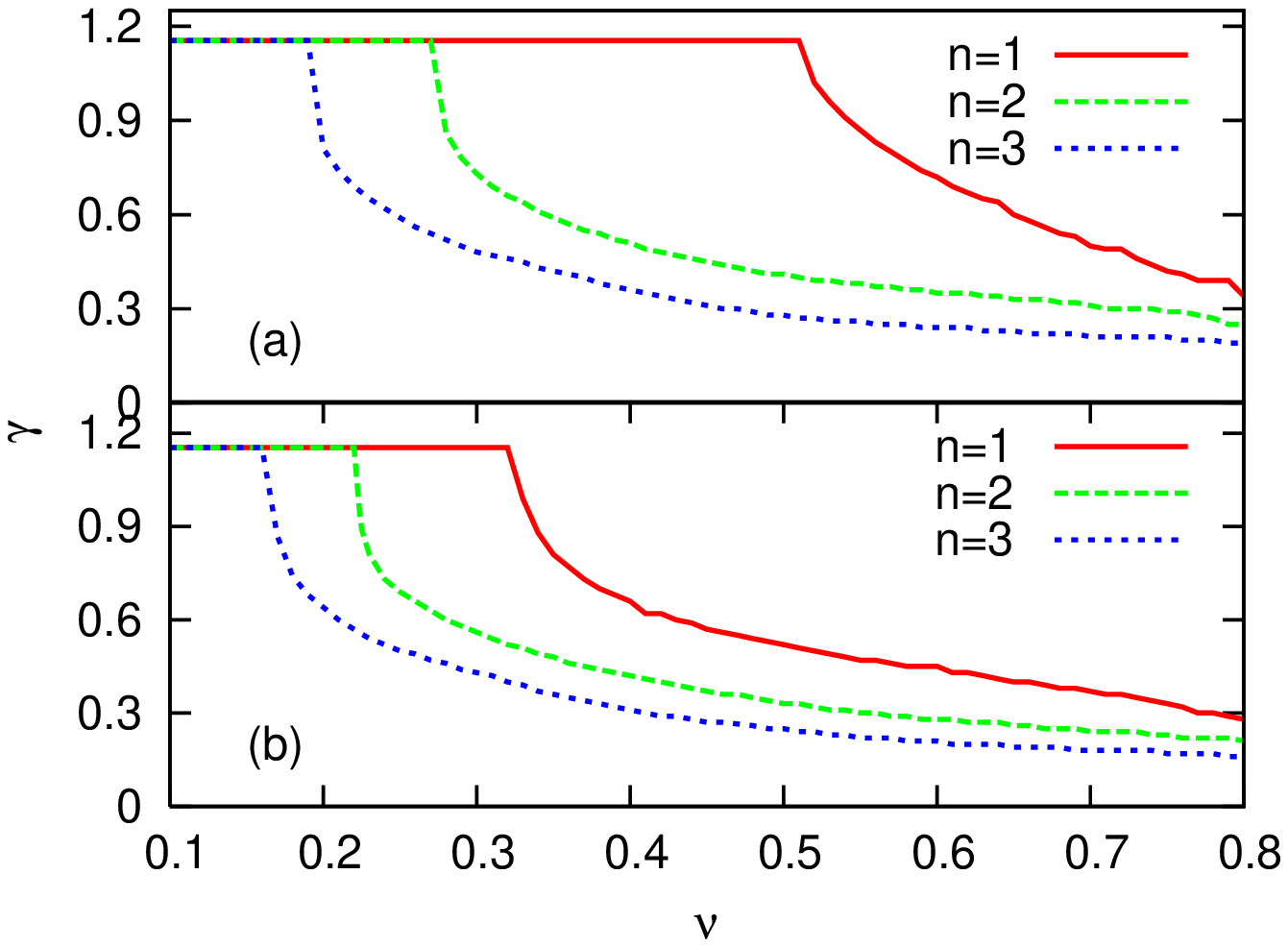}
\caption{(Color Online) Optimal crystal structure parameter
$\gamma(\nu)$ for the ground state of (a) graphene and (b) bilayer system
at $d=0$ (bottom) for different Landau level indices. Recall that
$\gamma=2/\sqrt{3}=1.15$ corresponds to a triangular lattice whereas
$\gamma\leq 0.5$ corresponds to highly anisotropic Wigner crystals or
striped states.}
\label{fig:gamma}
\end{figure}

For $n\ge1$, the equivalence between graphene and a bilayer quantum
Hall system at layer separation $d=0$ breaks down since the
single-particle wave functions are different. Figures
\ref{fig:ehfnun1}-\ref{fig:ehfnun3} show the phase diagram as a
function of partial filling factor $\nu$ for Landau levels $n$=1-3,
respectively. The phase diagram of $n=1$ Landau level is
qualitatively similar to $n=0$ Landau level. For $n\ge2$ we find
that the ground state for graphene is an isotropic crystal when the
partial filling factor $\nu$ is small, whereas it is an anisotropic
crystal when $\nu$ is sufficiently large. For intermediate values of
$\nu$, we find that the ground state is a bubble crystal with
$N_e=2$ or $N_e=3$. For example, for $n=2$ Landau level in graphene,
the triangular Wigner crystal is stable for $\nu\leq 0.28$, the
bubble state with two electrons $N_e=2$ is stable for
$0.28\leq\nu\leq 0.43$, whereas for $\nu\geq 0.43$ an anisotropic
Wigner crystal has the lowest energy. We can also see that the
critical values of $\nu$ at which transitions from a triangular
Wigner crystal to a bubble state to an anisotropic Wigner crystal
take place are systematically higher than corresponding values for a
bilayer system at $d=0$. Figure\ \ref{fig:gamma} shows optimal value
of $\gamma(\nu)$ for the ground state crystal structure for
different Landau level indices. We see that the transition from an
isotropic Wigner crystal ($\gamma=2/\sqrt{3}=1.15$) to an
anisotropic Wigner crystal in graphene (a) occurs at higher values
of $\nu$ than the corresponding values in bilayer systems (b).


\begin{figure}[htb]
\includegraphics[width=8cm]{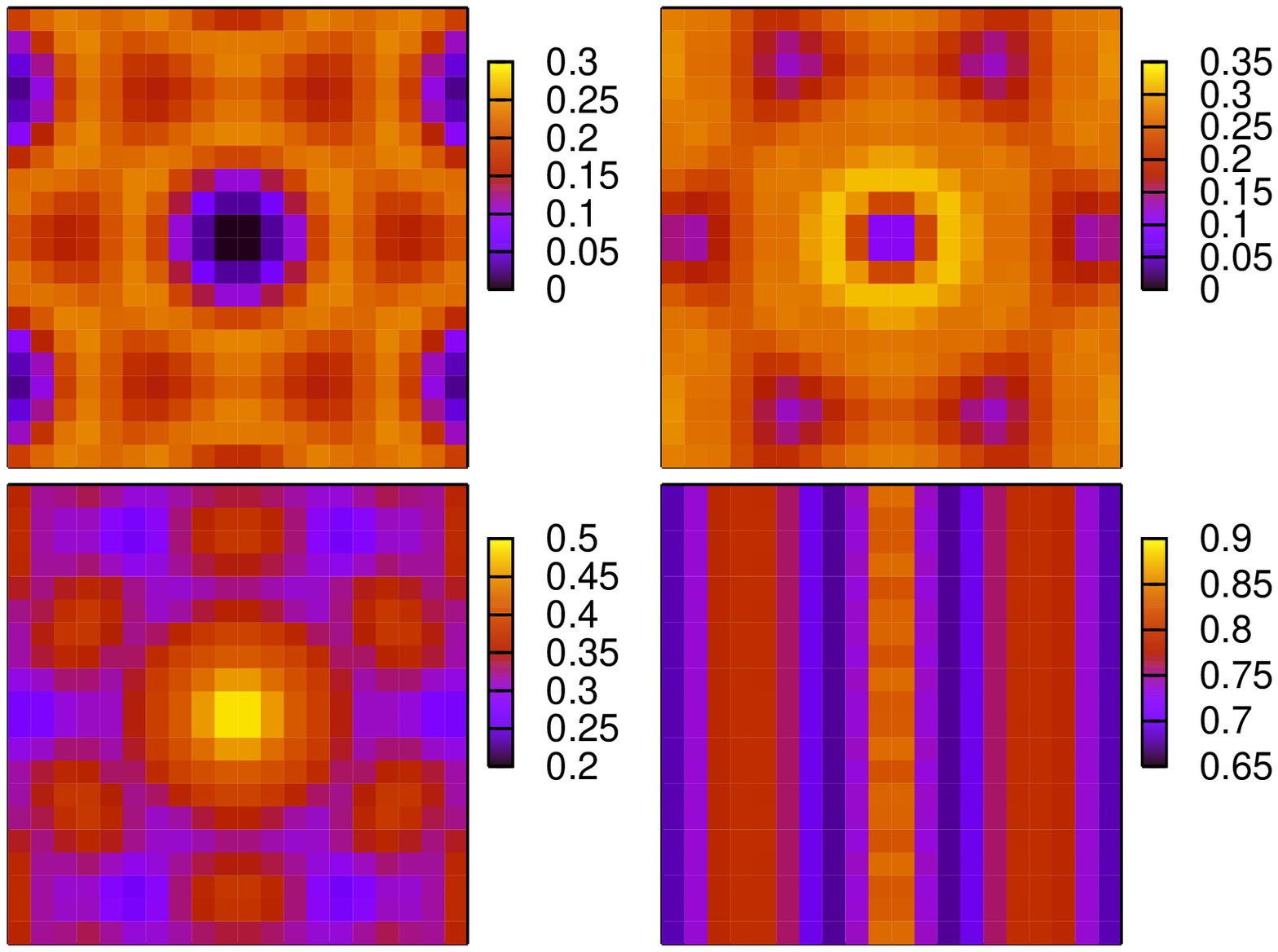}
\caption{(Color Online) A density plot for the ground state electron
density in graphene with Landau level $n=3$. Shown here are a
triangular Wigner crystal at $\nu=0.18$ (top left), a bubble state with
$N_e=2$ at $\nu=0.25$ (top right), a bubble state with $N_e=3$ that occurs
at $\nu=0.35$ (bottom left), and the anisotropic Wigner crystal
state at $\nu=0.75$ (bottom right).}
\label{fig:rhon3}
\end{figure}

Figure\ \ref{fig:rhon3} shows the (dimensionless) real-space electron
density profile for the
$n=3$ Landau level when the system is in the isotropic Wigner
crystal state ($\nu=0.18$), in the bubble crystal state with $N_e=2$
($\nu=0.25$) and with $N_e=3$ ($\nu=0.35$), and in the anisotropic
Wigner crystal state ($\nu=0.75$). We see that the density profile
is different from that in the lowest Landau level, due to different
form factors.
It is instructive to compare density profiles of the anisotropic
Wigner crystal in the $n=3$ Landau level and $n=0$ Landau level
(Fig.\ \ref{fig:rhon0nu0.75}). We see that both resemble a
(quasi)-uniform striped states, although the $n=3$ density profile
also shows the existence of uniform stripes between the modulated
ones.

\begin{figure}[thpb]
\includegraphics[width=13cm]{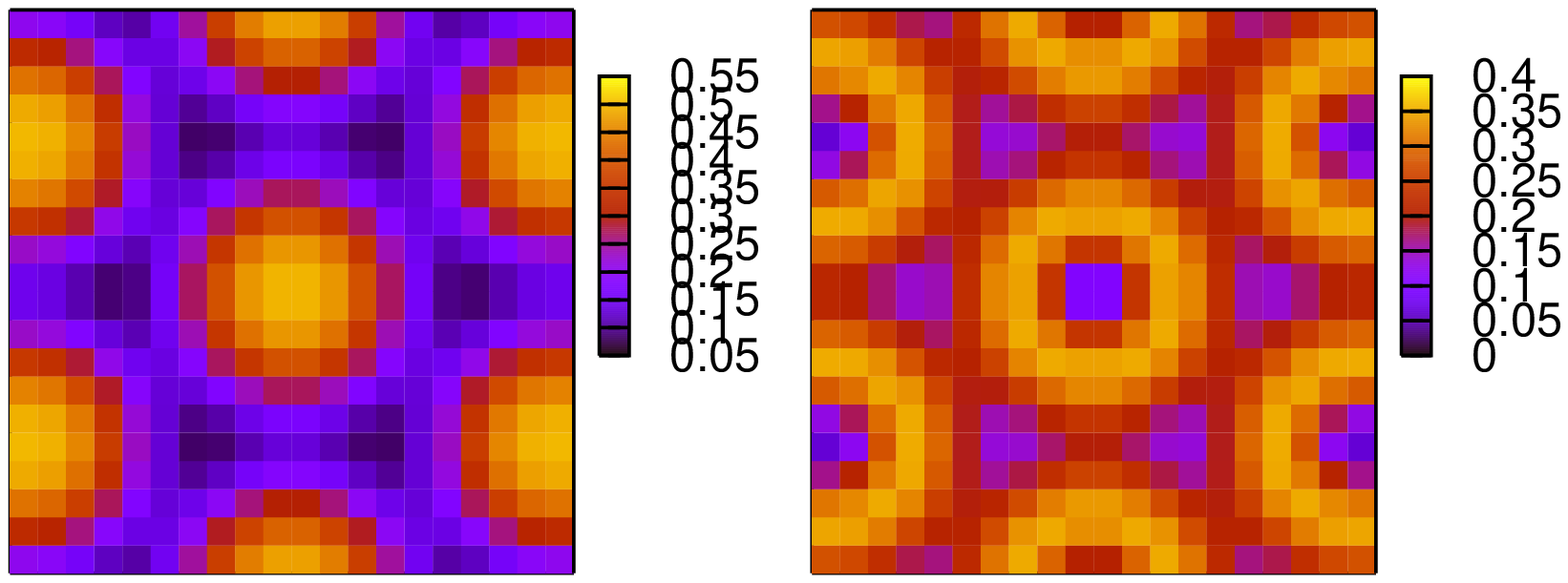}
\caption{(Color Online) A density plot of the ground state electron
density profile in graphene (left) and bilayer system at $d=0$
(right) at partial filling factor $\nu=0.25$ in the $n=1$ Landau
level. Both systems have identical density matrices $\rho_n(\bQ)$.
However, the difference in the form factors in Eq.\ (\ref{eq:rhor})
leads to different real space densities.} \label{fig:rhonu0.25}
\end{figure}

We end this section by comparing the electron density profiles in
graphene and bilayer system with $d=0$ at partial filling $\nu=0.25$
in the $n=1$ Landau level in Fig.\ \ref{fig:rhonu0.25}. Recall that
in the lowest Landau level, these density profiles are identical
(Fig\ \ref{fig:rhon0nu0.25}). For both systems, the self-consistent
solutions for the density matrix $\rho_n^{\sigma\sigma'}(\bq)$ are
identical; however, for $n\geq 1$, due to the differences in form
factors, the resultant real-space electron density profiles are
different.

\section{Discussion and Conclusions}
\label{sec:end}

In this paper, we have systematically studied the ground state of
graphene in the presence of a strong magnetic field, focusing on
broken symmetry states with intervalley coherence, using the
Hartree-Fock mean-field analysis. We have ignored the inter-Landau
level transitions (since their inclusion does not qualitatively change the 
phase diagram) and considered spin-split Landau levels. We have focused
only states with inter-valley coherence because, due to the
isotropic Coulomb interaction,energy of a state without coherence is always 
larger than the energy of a state with intervalley coherence. Thus, in 
experiments, these levels will correspond to total filling factors 
$0<\nu_T<1$ (lowest
subband of the $n=0$ level), $2<\nu_T<3$ (lowest subband of the
$n=1$ level), $6<\nu_T< 7$ (lowest subband of the $n=2$ level). Our 
calculations, naturally, have also not taken into account competing 
(Laughlin) fluid states with uniform density. Such fluid states will have 
lower energy at special 
filling factors, for example, $\nu=1/3$; however, our results likely
represent the true ground state of the system at generic filling
factors and in high Landau levels (where, in a conventional 2DEG, 
Hartree-Fock solutions are reliable).

Our analysis found that, since the kinetic energy of graphene's
linearly dispersing carriers is quenched in the presence of a
magnetic field, graphene is qualitatively similar to a conventional
2DEG with a quantum degree of freedom (bilayer quantum Hall system
at $d=0$ or a single layer system with vanishing Zeeman coupling).
We showed that the (mean-field) ground state of graphene is evolves
from a triangular Wigner crystal to an anisotropic Wigner crystal
(striped state) as the partial filling $\nu$ in a given Landau level
is increased. We also showed that for Landau level indices $n\geq
2$, a bubble crystal state with two electrons per unit cell, $N_e=2$
occurs at intermediate values of $\nu$.\ \cite{Koulakov96}
We have compared our results with mean-field results for a
corresponding bilayer system at $d=0$ (where the Coulomb interaction
becomes isotropic in the pseudospin space). Our findings indicate
that different form factors for electrons in graphene systematically
shift the critical values of $\nu$ at which the phase transitions
occur to higher values, {\it thus expanding the region of stability
for the triangular Wigner crystal and bubble crystal states,
compared to their counterparts in bilayer systems}.


These broken symmetry states in graphene are open to more probes
than the conventional 2DEG. Our analysis predicts that anisotropies
in the longitudinal resistance will be observed in high Landau
levels (similar to those observed in quantum Hall systems) and will
provide a signature of highly anisotropic Wigner crystal (striped)
states. In addition, since the 2DEG in graphene is literally at the
surface, the electron density modulations can be directly probed, by
scanning tunneling microscopy, and may provide a {\it direct}
evidence of electronic crystal states. A weak random disorder will, in 
general, pin these crystals and destroy the true long-range order. However, 
since it does not prefer one crystal structure over another, the disorder 
will not change the phase diagram qualitatively. Direct experimental
observation of charge and current density distribution in quantum
Hall systems is an outstanding problem; observation of isotropic
Wigner crystal and striped states in graphene will improve our
understanding of local structure of quantum Hall states.

\section{Acknowledgments}
It is a pleasure to thank Herb A. Fertig, Sasha Balatsky, and Allan
MacDonald for helpful discussions.

\bibliographystyle{apsrev}
\bibliography{references2}

\end{document}